\pdfoutput=1
\documentclass[prl, twocolumn,superscriptaddress]{revtex4-1}
\usepackage[colorlinks,
            linkcolor=blue,
            anchorcolor=blue,
            citecolor=blue]{hyperref}

\usepackage{times}
\usepackage{graphicx}
\usepackage{amssymb}
\usepackage{amsmath}
\usepackage{bm}
\usepackage{color}
\usepackage{float}
\usepackage{setspace}

\usepackage{subfigure}

\usepackage{siunitx}
\usepackage[a4paper, left=1.7cm, right=1.7cm, top=2cm, bottom=3cm]{geometry}

\usepackage{orcidlink}

\begin{document}

\title{Third-order orbital corner state and its realization in acoustic crystals}

\author{Jiyu Wang${~\orcidlink{0000-0002-7285-166X}}$}

\affiliation{Department of Physics,  Xiamen University, Xiamen 361005, China} 

\author{Ying Chen}

\affiliation{Department of Physics, College of Information Science and Engineering, Huaqiao University, Xiamen 361021, China} 

\author{Xiancong Lu${~\orcidlink{0000-0001-9863-7015}}$}
\email[Corresponding author: ]{xlu@xmu.edu.cn}
\affiliation{Department of Physics,  Xiamen University, Xiamen 361005, China}

\begin{abstract}
Three dimensional (3D) third-order topological insulators (TIs) have zero-dimensional (0D) corner states, which are three dimensions lower than bulk. Here we investigate the third-order TIs on breathing pyrochlore lattices with $p$-orbital freedom. 
The tight-binding Hamiltonian is derived for the $p$-orbital model, in which we find that the two orthogonal $\pi$-type (transverse) hoppings 
are the key to open a band gap and obtain higher-order topological corner states. 
We introduce the $Z_4$ berry phase to characterize the bulk topology and analysis the phase diagram.
The corner states, demonstrated in a finite structure of a regular tetrahedron, exhibit rich 3D orbital configurations. Furthermore, we design an acoustic system to introduce the necessary $\pi$-type hopping and successfully observe the orbital corner states. 
Our work extends topological orbital corner states to third-order, which enriches the contents of orbital physics and may lead to applications in novel topological acoustic devices.

\end{abstract}

\date{\today}
\maketitle

\section*{\romannumeral1. Introduction}

Topological insulators (TIs), well known for their bulk–boundary correspondence, are new states of quantum matter 
with protected gapless edge or surface states \cite{kane2005z,kane2005quantum,bernevig2006quantum,konig2007quantum,fu2007topological,moore2010birth,hasan2010colloquium,qi2011topological}. 
In recent years, the concept of higher-order topological insulators (HOTIs) have been put forward, which 
host gapless hinge or corner states at lower dimensions instead of edge or surface states \cite{benalcazar2017quantized,benalcazar2017electric,song2017d,langbehn2017reflection,geier2018second,schindler2018higher}. The HOTIs are extensively studied in two-dimensional (2D) and three-dimensional (3D) systems \cite{kim2020recent,xie2021higher,yang2024higher}, and have even yielded applications such as corner state lasers \cite{zhang2020low,kim2020multipolar,wu2023higher}.

Breathing pyrochlore lattices, the 3D extension of breathing kagome lattices, have been experimentally realized in A-site ordered spinel oxides \cite{okamoto2013breathing}. Various phenomena have been revealed, such as the magnetic phase transitions
\cite{tanaka2014novel}, the unusual octupolar paramagnet \cite{rau2016anisotropic}, the origin of geometrical frustration \cite{pokharel2020cluster} and the spin resonances in the magnetically ordered state \cite{he2021neutron}. 
Regarding the theoretical works about quantum magnetism on breathing pyrochlore lattices, Weyl magnons have been proposed \cite{li2016weyl}, 
magnetic phases in the presence of competing interactions have been studies \cite{iqbal2019quantum}, and a spin liquid described by rank-2 $U(1)$ gauge theory has been demonstrated \cite{yan2020rank}.
 Recently, the higher-order topology on breathing pyrochlore lattices also attracts many attentions: 
 the topology is characterized by the quantized polarization \cite{ezawa2018higher,ezawa2018higher2}; a generic recipe for exactly solvable "boundary" states is proposed \cite{kunst2018lattice};
 the corner states are successfully observed in the acoustic metamaterials \cite{weiner2020demonstration}. 
Despite these exciting discoveries, the study of orbital physics in breathing pyrochlore lattices has been largely overlooked.

Over the past few decades, orbital physics has been extensively studied in transition metal compounds \cite{tokura2000orbital, khomskii2022orbital} and ultracold atomic systems in optical lattices \cite{lewenstein2011orbital,li2016physics}. 
When occupying $p$-bands of optical lattices, the orbital order and Mott-insulating state have been 
proposed for the spinless fermions \cite{zhao2008orbital,lu2009dispersive}, and the orbital superfluidity has been 
demonstrated for bosons \cite{wirth2011evidence,soltan2012quantum}. 
In recent years, topological properties of $p$-orbital systems have received significant research attention \cite{milicevic2017orbital,gao2023orbital,schulz2022photonic,lu2020orbital,zhang2023realization,gao2024acoustic,wu2024observation,bongiovanni2024p-orbital,chen2022observation,wang2022valley,liu2024monoatomic,li2024disentangled}. 
The topological edge states of $p$-orbital were observed in photonic crystals \cite{milicevic2017orbital} and acoustic resonator chains \cite{gao2023orbital}. 
The $p$-orbital modes were also utilized to generate synthetic magnetic flux in a quadrupole topological photonic lattice \cite{schulz2022photonic}. Moreover, the second-order $p$-orbital corner states have been theoretically proposed in (2D) breathing kagome lattice \cite{lu2020orbital}, 
which are successfully realized using photonic systems \cite{zhang2023realization}. 
However, previous studies on $p$-orbital HOTIs, such as those in Refs. \cite{lu2020orbital,zhang2023realization,gao2024acoustic,wu2024observation,bongiovanni2024p-orbital,chen2022observation}, have been limited to two-dimensional or quasi-2D systems, typically involving two $p$-orbitals ($p_x$ and $p_y$). 

In this paper, we address the unexplored domain of higher-order band topology in 3D $p$-orbital systems, filling a critical gap in the field. 
By incorporating all hopping processes of three $p$-orbitals in 3D space \cite{chen2020simulating,chen2022emergent,klosinski2023topology,luo2023bosonic}, we construct the proper tight-binding model for the $p$-orbital breathing pyrochlore lattice and explore the intriguing interplay between third-order band topology and the orbital degree of freedom.  
Furthermore, guided by the predictions of orbital corner states from the tight-binding model, we design the structures of the acoustic crystals to ensure sufficient $\pi$-type hopping strength. This enabled the observation of third-order orbital corner states within a well-defined gap, representing the realization of such states in a 3D acoustic system.

The paper is organized as follows:
In Sec. \hyperlink{s2}{\uppercase\expandafter{\romannumeral2}}, we introduce the tight-binding Hamiltonian for $p$-orbital model, where we emphasize the importance of two orthogonal $\pi$-type hopping in 3D lattice. 
In Sec. \hyperlink{s3}{\uppercase\expandafter{\romannumeral3}}, we study the bulk topological properties of $p$-orbital breathing pyrochlore lattice, including the band structure and the $Z_4$ berry phase.
In Sec. \hyperlink{s4}{\uppercase\expandafter{\romannumeral4}}, we construct a four-layer regular tetrahedral finite structure to demonstrate the zero-energy corner states as well as their rich orbital configurations. 
In Sec. \hyperlink{s5}{\uppercase\expandafter{\romannumeral5}}, the $p$-orbital corner states are realized in acoustic system by using numerical simulation. Finally, we provide a summary in Sec. \hyperlink{s6}{\uppercase\expandafter{\romannumeral6}}

\hypertarget{s2}{}
\section*{\romannumeral2. Tight-binding Hamiltonian}


\begin{figure*}[t]
\centering
\includegraphics[width=0.99\linewidth]{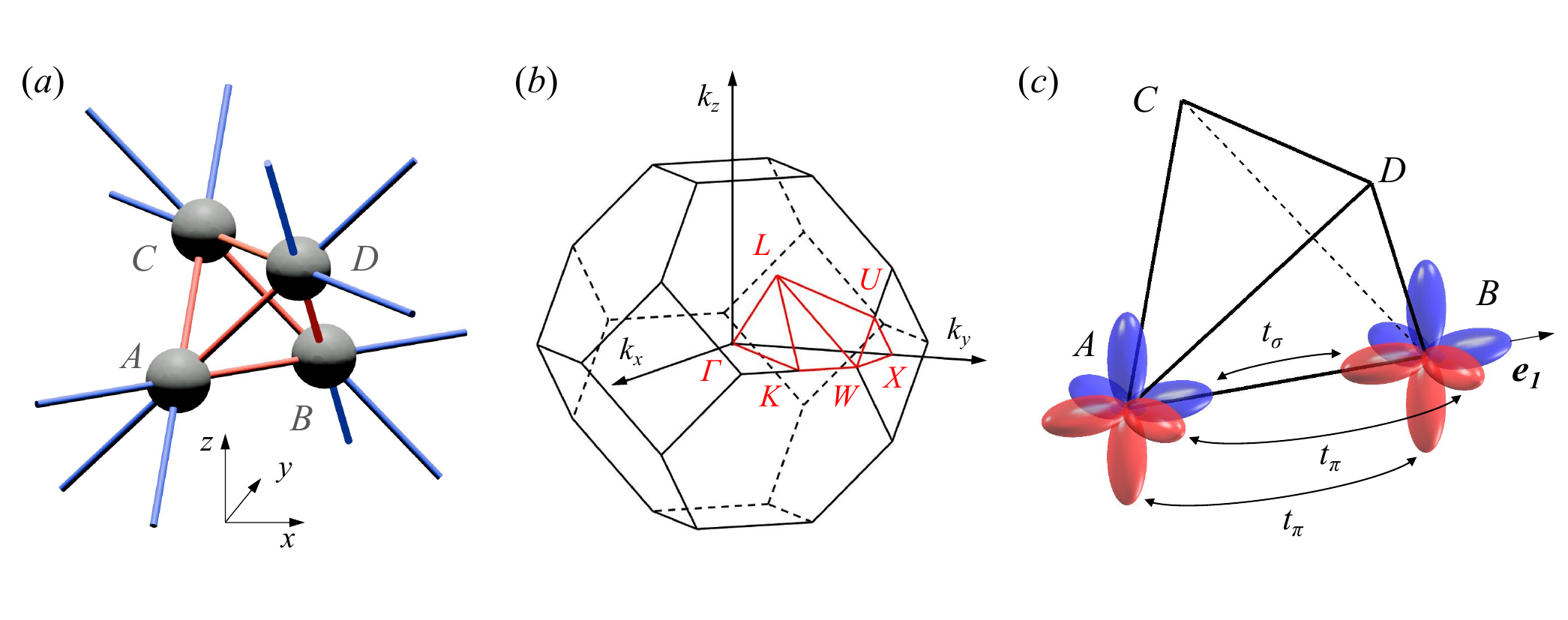}
\caption{(a) The structure of breathing pyrochlore lattices in real space. Gray solid spheres represent four sites in a unit cell, which make up a regular tetrahedron. The red and blue cylinders indicate the hoppings inside and between unit cells. (b) The first Brillouin zone with high symmetry points ($\Gamma,X,W,U,L,K$) in $k$ space. (c) The $\sigma$ and $\pi$ types of hoppings for $p$ orbitals. For 3D lattices, two orthogonal $\pi$ hoppings should be taken into account. The red and blue ellipsoids represent the signs ($+$ and $-$) of two lobes of a $p$ orbital.}
\label{f1}
\end{figure*}

The structure of breathing pyrochlore lattices is shown in Fig. \ref{f1}(a), 
where $A$, $B$, $C$ and $D$ denote four sites in a unit cell. The $t_1$ and $t_2$ are intracell and intercell hopping amplitudes respectively where breathing hopping ($t_1\neq t_2$) 
leads to the HOTIs \cite{ezawa2018higher}. 
We then construct the tight-binding Hamiltonian of breathing pyrochlore lattice with $p$ orbitals.
For a 3D system, each lattice site allows for three $p$ orbitals: 
the $p_x , p_y$ and $p_z$ orbitals. There are basically two types of hoppings, \textit{i.e.,} the
$\sigma$-type (longitudinal) and $\pi$-type (transverse), whose amplitudes are represented by $t_{\sigma}$ 
and $t_\pi$ respectively \cite{liu2006atomic,wu2008p,milicevic2019type,lu2020orbital,sun2023montage}. 
Note that only one $\pi$-type hopping needs to be considered for the 2D $p$-orbital models studied in previous works \cite{liu2006atomic,wu2008p,milicevic2019type,lu2020orbital,sun2023montage}. However, for 3D models, one needs to take into account two orthogonal $\pi$-type hoppings, to fully capture the $\pi$-type hopping process; see Fig. \ref{f1}(c). 

For the $\sigma$-type (longitudinal) hopping, we define six vectors along the hopping directions between 
two nearest-neighboring lattice sites,

\begin{equation}
\begin{aligned}
\centering
&\bm{e}_1= (1 , 1 , 0), \quad
\bm{e}_2= (0 , 1 , 1), \quad
\bm{e}_3= (1 , 0 , 1),\\
&\bm{e}_4= (-1 , 0 , 1), 
\bm{e}_5= (0 , -1 , 1), 
\bm{e}_6= (1 , -1 , 0).
\label{q1}
\end{aligned}
\end{equation}

The $p$ orbitals can be projected along these hopping directions of lattices. The six projection operators $p^\sigma_i$$(i=1,2,\cdots, 6)$ can be written as

\begin{equation}
\begin{aligned}
\centering
&{p}_{1,6}^{\sigma}=\bm{e}_{1,6} \cdot \bm{p}=p_x \pm p_y , \\
&{p}_{2,5}^{\sigma}=\bm{e}_{2,5} \cdot \bm{p}=\pm p_y + p_z , \\
&{p}_{3,4}^{\sigma}=\bm{e}_{3,4} \cdot \bm{p}=\pm p_x + p_z , \\
\end{aligned}
\end{equation}

in which the $p$-orbital operator $\bm{p}$ is defined on the $p_x$ , $p_y$ and $p_z$ basis: $\bm{p}=(p_x, p_y, p_z)^T$. 

For the $\pi$-type (transverse) hopping, we define two sets of vectors $\bm{m}_i$ and $\bm{n}_i$ $(i=1,2,\cdots, 6)$ 
which are perpendicular to the hopping directions:

\begin{equation}
\begin{aligned}
\centering
&\bm{m}_1= (1 , -1 , 0), 
\bm{m}_2= (0 , 1 , -1), 
\bm{m}_3= (1 , 0 , -1),\\
&\bm{m}_4= (1 , 0 , 1), \quad
\bm{m}_5= (0 , 1 , 1), \quad
\bm{m}_6= (1 , 1 , 0).
\end{aligned}
\end{equation}

\begin{equation}
\begin{aligned}
\centering
&\bm{n}_1= (0 , 0 , \sqrt{2}), \quad
\bm{n}_2= (\sqrt{2} , 0 , 0), \quad
\bm{n}_3= (0 , \sqrt{2} , 0),\\
&\bm{n}_4= (0 , -\sqrt{2} , 0), 
\bm{n}_5= (-\sqrt{2} , 0 , 0), 
\bm{n}_6= (0 , 0 , -\sqrt{2}).
\end{aligned}
\end{equation}

It is worth noting that the vectors $\bm{e}_i$, $\bm{m}_i$ and $\bm{n}_i$ are pairwise orthogonal to each other. 
Then, two sets of orthogonal projection operators ($p^{\pi 1}_i$  and $p^{\pi 2}_i$) perpendicular to the hopping direction are given by

\begin{equation}
\begin{aligned}
\centering
&{p}_{1,6}^{\pi1}=\bm{m}_{1,6} \cdot \bm{p}=p_x \mp p_y , 
&{p}_{2,5}^{\pi1}=\bm{m}_{2,5} \cdot \bm{p}=p_y \mp p_z , \\
&{p}_{3,4}^{\pi1}=\bm{m}_{3,4} \cdot \bm{p}=p_x \mp p_z , 
&{p}_{1,6}^{\pi2}=\bm{n}_{1,6} \cdot \bm{p}=\pm{\sqrt{2}}p_z ,\\
&{p}_{2,5}^{\pi2}=\bm{n}_{2,5} \cdot \bm{p}=\pm{\sqrt{2}}p_x,
&{p}_{3,4}^{\pi2}=\bm{n}_{3,4} \cdot \bm{p}=\pm{\sqrt{2}}p_y.
\end{aligned}
\end{equation}

\begin{figure*}[t]
\centering
\includegraphics[width=0.99\linewidth]{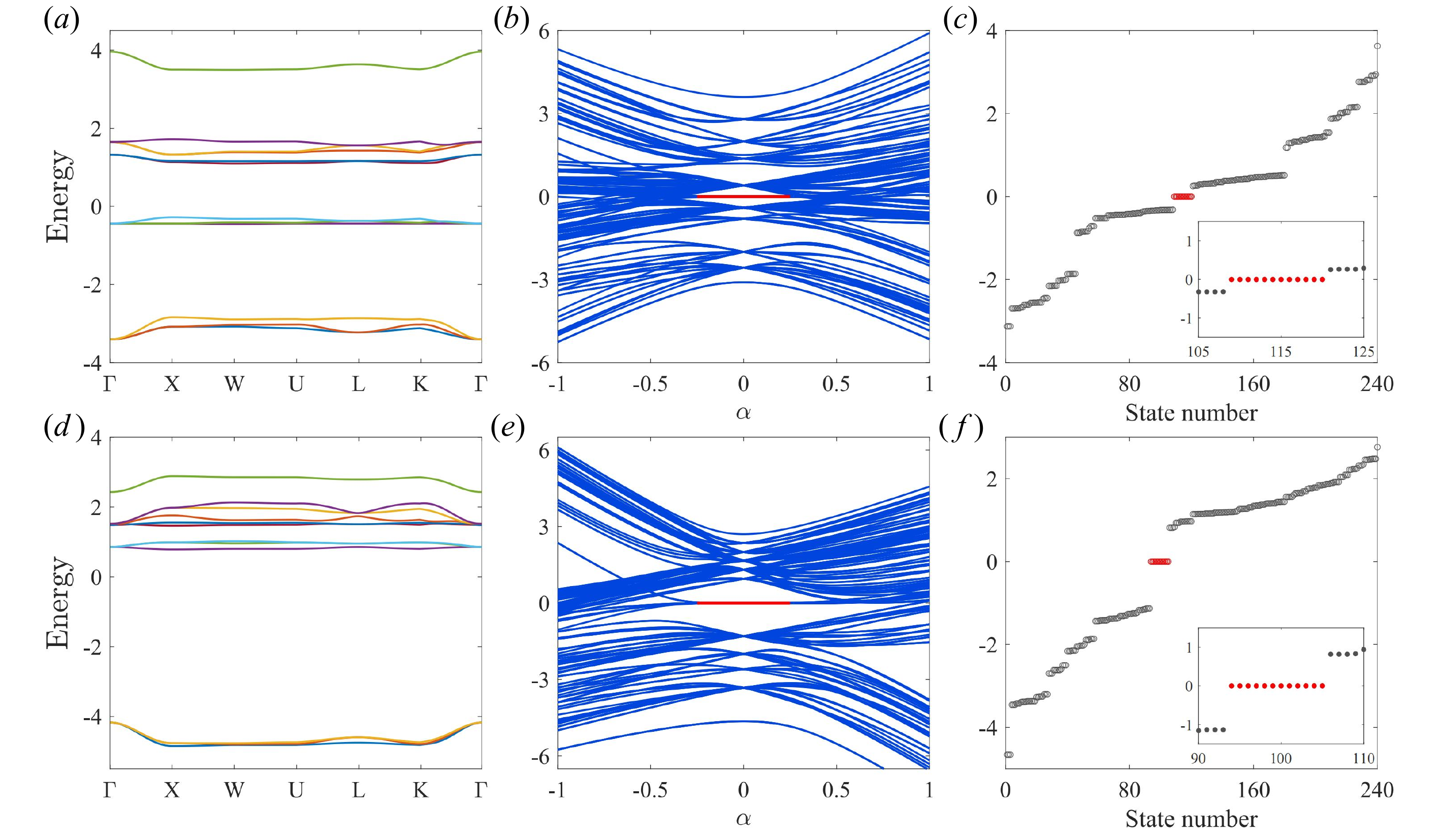}
\caption{(a), (d) Band structures of $p$-orbital breathing pyrochlore lattice in k-space. The values of hopping amplitude are (a) $\alpha=0.1, \beta=0.2$; six bands under the zero-energy gap; (d) $\alpha=-0.1, \beta=0.65$; three bands under the zero-energy gap; (b), (c), (e), (f) The energy spectrum of a four-layer ($L=4$) tetrahedron-shaped finite structure of $p$-orbital breathing pyrocholre lattice. The energy dispersion as a function of $\alpha$ when (b) $\beta=0.2$ and (e) $\beta=0.65$. 
The horizontal red lines indicate the orbital corner states with $E=0$. 
The energy as a function of state number when (c) $\alpha=0.1$ and $\beta=0.2$, (f) $\alpha=-0.1$ and $\beta=0.65$.
The 12 degenerate corner states are magnified in the inset figures in (c) and (f).}
\label{f2}
\end{figure*}

With these definitions, we obtain the real-space Hamiltonian for the $p$-orbital breathing pyrochlore lattices,

\begin{equation}
\begin{aligned}
\centering
H=-\sum_{\bm{r}}\Big[&t_{1\sigma}
   (a_{\bm{r},1}^{\sigma\dagger}b_{\bm{r},1}^{\sigma}
   +a_{\bm{r},2}^{\sigma\dagger}c_{\bm{r},2}^{\sigma}
   +a_{\bm{r},3}^{\sigma\dagger}d_{\bm{r},3}^{\sigma}\\
  &+b_{\bm{r},4}^{\sigma\dagger}c_{\bm{r},4}^{\sigma}
   +b_{\bm{r},5}^{\sigma\dagger}d_{\bm{r},5}^{\sigma}
   +c_{\bm{r},6}^{\sigma\dagger}d_{\bm{r},6}^{\sigma})\\
+t_{2\sigma}&  
   (a_{\bm{r}+\bm{e}_1,1}^{\sigma\dagger}b_{\bm{r},1}^{\sigma}
   +a_{\bm{r}+\bm{e}_2,2}^{\sigma\dagger}c_{\bm{r},2}^{\sigma}
   +a_{\bm{r}+\bm{e}_3,3}^{\sigma\dagger}d_{\bm{r},3}^{\sigma}\\
  &+b_{\bm{r}+\bm{e}_4,4}^{\sigma\dagger}c_{\bm{r},4}^{\sigma}
   +b_{\bm{r}+\bm{e}_5,5}^{\sigma\dagger}d_{\bm{r},5}^{\sigma}
   +c_{\bm{r}+\bm{e}_6,6}^{\sigma\dagger}d_{\bm{r},6}^{\sigma})\\
+t_{1\pi}&
   (a_{\bm{r},1}^{\pi\dagger}b_{\bm{r},1}^{\pi}
   +a_{\bm{r},2}^{\pi\dagger}c_{\bm{r},2}^{\pi}
   +a_{\bm{r},3}^{\pi\dagger}d_{\bm{r},3}^{\pi}\\
  &+b_{\bm{r},4}^{\pi\dagger}c_{\bm{r},4}^{\pi}
   +b_{\bm{r},5}^{\pi\dagger}d_{\bm{r},5}^{\pi}
   +c_{\bm{r},6}^{\pi\dagger}d_{\bm{r},6}^{\pi})\\
+t_{2\pi}& 
  (a_{\bm{r}+\bm{e}_1,1}^{\pi\dagger}b_{\bm{r},1}^{\pi}
  +a_{\bm{r}+\bm{e}_2,2}^{\pi\dagger}c_{\bm{r},2}^{\pi}
  +a_{\bm{r}+\bm{e}_3,3}^{\pi\dagger}d_{\bm{r},3}^{\pi}\\
 &+b_{\bm{r}+\bm{e}_4,4}^{\pi\dagger}c_{\bm{r},4}^{\pi}
  +b_{\bm{r}+\bm{e}_5,5}^{\pi\dagger}d_{\bm{r},5}^{\pi}
  +c_{\bm{r}+\bm{e}_6,6}^{\pi\dagger}d_{\bm{r},6}^{\pi})
  \Big] \\
  + H.c.
\label{q6}
\end{aligned}
\end{equation}
in which each $\pi$-type hopping term (see the last 12 terms in Eq. (\ref{q6})) includes two identical types of hopping ($\pi_1$ and $\pi_2$), 
for example,  $a_{\bm{r},1}^{\pi\dagger}b_{\bm{r},1}^{\pi}=a_{\bm{r},1}^{\pi_1\dagger}b_{\bm{r},1}^{\pi_1}+a_{\bm{r},1}^{\pi_2\dagger}b_{\bm{r},1}^{\pi_2}$. 
$t_{1\sigma}$ and $t_{1\pi}$ ($t_{2\sigma}$ and $t_{2\pi}$) represent the amplitudes of the $\sigma$ and $\pi$ types of hopping inside (between) unit cells. $a^\sigma_{\bm{r}}$($a^\pi_{\bm{r}}$), $b^\sigma_{\bm{r}}$($b^\pi_{\bm{r}}$), 
$c^\sigma_{\bm{r}}$($c^\pi_{\bm{r}}$), and $d^\sigma_{\bm{r}}$($d^\pi_{\bm{r}}$) are the $\sigma$ ($\pi$) type of annihilating
projection operators at the sites $A$, $B$, $C$, and $D$, respectively, in
the unit cell located at position $\bm{r}$.
The right-side subscript $i$ in operator $a^\sigma_{\bm{r},i}$ ($a^\pi_{\bm{r},i}$) indicates that the projection is along the direction 
$\bm{e}_i$.

After making Fourier transformation to Eq. (\ref{q6}) and introducing a 12-component spinor 
\begin{small} 
$\psi=[a_{\bm{k},x}, a_{\bm{k},y}, a_{\bm{k},z}, b_{\bm{k},x}, b_{\bm{k},y}, b_{\bm{k},z}, c_{\bm{k},x}, c_{\bm{k},y}, c_{\bm{k},z}, d_{\bm{k},x}, d_{\bm{k},y}, d_{\bm{k},z}]^T,$
\end{small} 
the Hamiltonian can be rewritten as $H=\sum_{\bm{k}}\psi^{\dagger}H(\bm{k})\psi$. The matrix $H(\bm{k})$ is derived as follows,
\begin{equation}
\centering
H(\bm{k})=-
\begin{pmatrix} 
0   && D_{1} && D_{2}&& D_{3}\\\\ 
D_{1}^{\dagger} && 0 && D_{4}&& D_{5}\\\\
D_{2}^{\dagger} && D_{4}^{\dagger} && 0 && D_{6} \\\\
D_{3}^{\dagger} && D_{5}^{\dagger} && D_{6}^{\dagger} && 0 
\label{q7}
\end{pmatrix}_{12\times12},
\end{equation}
in which the matrices are given by

\begin{figure*}[t]
\centering
\includegraphics[width=0.99\linewidth]{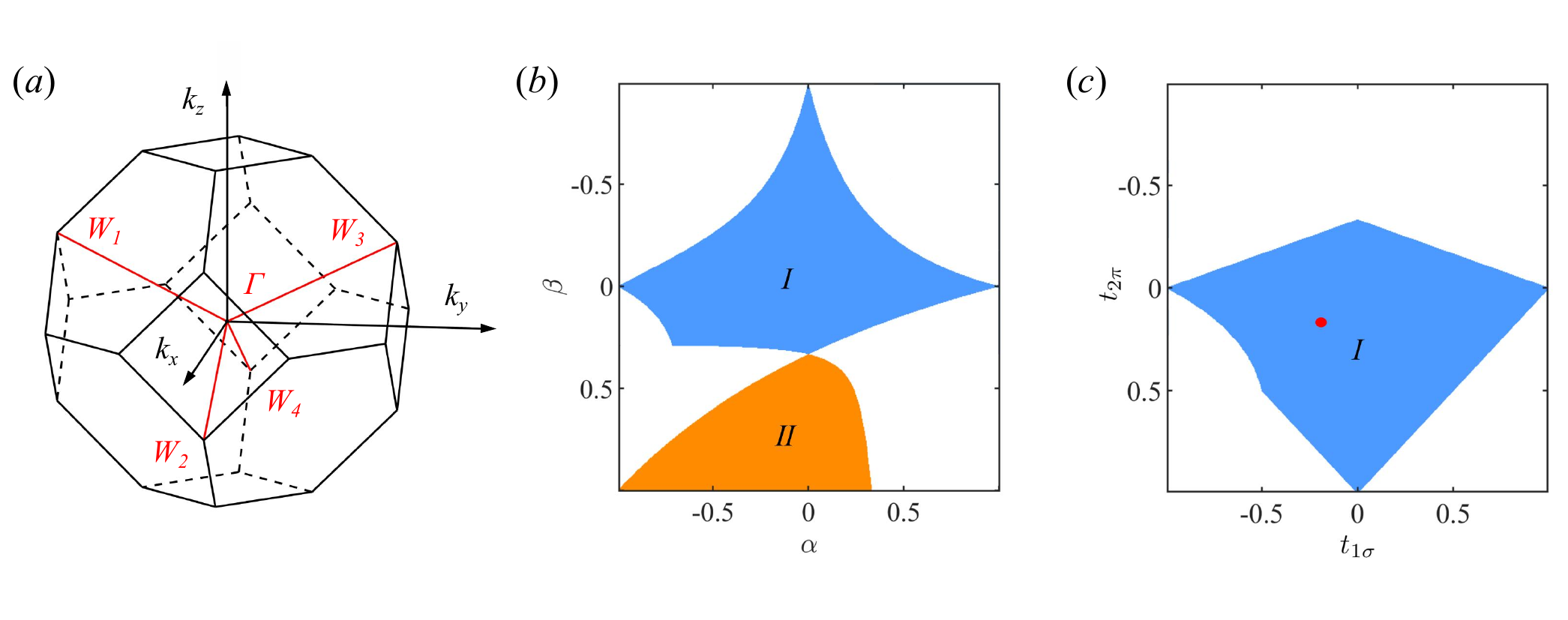}
\caption{Bulk topological properties of $p$-orbital breathing pyrochlore lattice. (a) Four integral paths (marked by red lines) in Brillouin zone: $P_1: W_1 \to \Gamma \to W_2$, $P_2: W_2 \to \Gamma \to W_3$, $P_3: W_3 \to \Gamma \to W_4$, $P_4: W_4 \to \Gamma \to W_1$. (b) and (c) Bulk topological phase diagram of $Z_4$ berry phase (mod $2\pi$) in the $\alpha-\beta$ and $t_{1\sigma}-t_{2\pi}$ planes. The orange area {\uppercase\expandafter{\romannumeral2}} corresponds to a value of 0.25 for the lowest three bands ($\nu$=3); The blue area {\uppercase\expandafter{\romannumeral1}} corresponds to a value of 0.5 for the lowest six bands ($\nu$=6). 
The red dot in the phase diagram (c) indicates the parameter values used in the acoustic simulation.}
\label{f3}
\end{figure*}

\begin{equation}
\begin{aligned}
\centering
D_1=
\begin{pmatrix} 
1 && 1 && 0\\
1 && 1 && 0\\
0 && 0 && 0 
\end{pmatrix}&f_{1\sigma}+
\begin{pmatrix} 
1  && -1 && 0\\
-1 && 1  && 0\\
0  && 0  && 2 
\end{pmatrix}f_{1\pi}
\end{aligned}
\end{equation}

\begin{equation}
\begin{aligned}
\centering
D_2=
\begin{pmatrix} 
0 && 0 && 0\\
0 && 1 && 1\\
0 && 1 && 1 
\end{pmatrix}&f_{2\sigma}+
\begin{pmatrix} 
2  && 0 && 0\\
0 && 1  && -1\\
0  && -1  && 1 
\end{pmatrix}f_{2\pi}
\end{aligned}
\end{equation}

\begin{equation}
\begin{aligned}
\centering
D_3=
\begin{pmatrix} 
1 && 0 && 1\\
0 && 0 && 0\\
1 && 0 && 1 
\end{pmatrix}&f_{3\sigma}+
\begin{pmatrix} 
1  && 0 && -1\\
0 && 2  && 0\\
-1  && 0  && 1 
\end{pmatrix}f_{3\pi}
\end{aligned}
\end{equation}
\begin{equation}
\begin{aligned}
\centering
D_4=
\begin{pmatrix} 
1 && 0 && -1\\
0 && 0 && 0\\
-1 && 0 && 1 
\end{pmatrix}&f_{4\sigma}+
\begin{pmatrix} 
1  && 0 && 1\\
0  && 2  && 0\\
1  && 0  && 1 
\end{pmatrix}f_{4\pi}
\end{aligned}
\end{equation}

\begin{equation}
\begin{aligned}
\centering
D_5=
\begin{pmatrix} 
0 && 0 && 0\\
0 && 1 && -1\\
0 && -1 && 1 
\end{pmatrix}&f_{5\sigma}+
\begin{pmatrix} 
2  && 0  && 0\\
0  && 1  && 1\\
0  && 1  && 1 
\end{pmatrix}f_{5\pi}
\end{aligned}
\end{equation}

\begin{equation}
\begin{aligned}
\centering
D_6=
\begin{pmatrix} 
1 && -1 && 0\\
-1 && 1 && 0\\
0 && 0 && 0 
\end{pmatrix}&f_{6\sigma}+
\begin{pmatrix} 
1  && 1  && 0\\
1  && 1  && 0\\
0  && 0  && 2 
\end{pmatrix}f_{6\pi}
\end{aligned}
\end{equation}

with $f_{i\sigma}=t_{1\sigma}+t_{2\sigma}e^{-i\bm{k}\cdot\bm{e_i}}$ and
$f_{i\pi}=t_{1\pi}+t_{2\pi}e^{-i\bm{k}\cdot\bm{e_i}}$.


\hypertarget{s3}{}
\section*{\romannumeral3. Bulk topological properties}

The Hamiltonian in Eq. (\ref{q6}) contains four hopping parameters: $t_{1\pi}$, $t_{1\sigma}$, $t_{2\pi}$, and $t_{2\sigma}$. In principle, one must consider the topologies in the four-parameters space. However, due to the specific situation in our simulation model, we can simplify our analysis to two parameters, allowing the results to be represented in a 2D phase diagram. We will discuss two scenarios here: the first one demonstrates richer content theoretically, while the second one corresponds to our subsequent acoustic simulation.
In systems such as photonic lattices, where intercell and intracell hoppings are tuned by adjusting the distance between two photonic cavities \cite{lu2020orbital}, the ratio between $\pi$-type and $\sigma$-type hoppings remains approximately constant, i.e., $t_{1\pi}/t_{1\sigma} \approx t_{2\pi}/t_{2\sigma}$.
We firstly introduce two additional parameters: $\alpha=t_{1\pi}/t_{2\pi}=t_{1\sigma}/t_{2\sigma}$ 
and $\beta=t_{1\pi}/t_{1\sigma} = t_{2\pi}/t_{2\sigma}$, and set $t_{2\sigma}$ as energy unit. 
This way allows us to conveniently discuss the topological phase diagram in the $\alpha-\beta$ plane which exhibits rich phase value.
For the acoustic systems to be discussed in Sec. \hyperlink{s5}{\uppercase\expandafter{\romannumeral5}}, 
the ratio between $\pi$-type and $\sigma$-type hoppings is 
no longer constant ($t_{1\pi}/t_{1\sigma} \neq t_{2\pi}/t_{2\sigma}$). 
However, 
as discussed in Sec. \hyperlink{s5}{\uppercase\expandafter{\romannumeral5}}, when the radius of the acoustic 
connecting cylinder is sufficiently small, the $\pi$-type hopping 
amplitude is negligibly small compared to the $\sigma$-type hopping. Therefore, given the small radius of the 
intracell cylinder in acoustic crystals, we can set $t_{1\pi}=0$ and focus on the topological properties in the $t_{1\sigma}-t_{2\pi}$ plane, 
using $t_{2\sigma}=-1$ as the energy unit. In the following, we examine the phase diagram 
in both $\alpha-\beta$ and $t_{1\sigma}-t_{2\pi}$ planes, which may be applicable to photonic and acoustic systems, respectively.


The Brillouin zone of breathing pyrochlore lattice with high symmetry points is shown in Fig. \ref{f1}(b), which forms a truncated octahedron. 
The energy bands are obtained by diagonalizing the $k$-space Hamiltonian in Eq. (\ref{q7}). There are four sites in a unit cell with three $p$ orbitals in each site, leading to 12 bands in this model. Here, we show the band structures in Fig. \ref{f2} where the zero energy appears in the band gap. The number of bands below the zero energy gap varies as the values of the parameters $\alpha$ and $\beta$ change. Typically, Fig. \ref{f2}(a) exhibits six bands under the zero energy gap with a relatively small value of $\beta$,
while Fig. \ref{f2}(d) exhibits three bands under the gap with a relatively large value of $\beta$. It is worth mentioning that the bands are degenerate when $\beta=1$, that is, all orbitals are isotropic. In this case, the band structure of our model go back to $s$-band model with three bands \cite{ezawa2018higher}, where the flat band is six-fold degenerate and the other two are three-fold degenerate.

$Z_Q$ berry phase is one of the symmetry protected topological invariants for HOTIs \cite{hatsugai2011zq,kariyado2018z,araki2020z}. Inspired by the $Z_3$ berry phase for the breathing kagome lattice \cite{lu2020orbital,wakao2020higher}, the $Z_4$ berry phase for the breathing square lattice \cite{li2020second} and recently $Z_6$ berry phase for distorted honeycomb lattice \cite{jiang2024higher},
we calculate the $Z_4$ berry phase in momentum space for the breathing pyrochlore lattice. The elements of the Berry connection matrix ($\nu \times \nu$) are


\begin{figure}[t]
\centering
\includegraphics[width=0.9\linewidth]{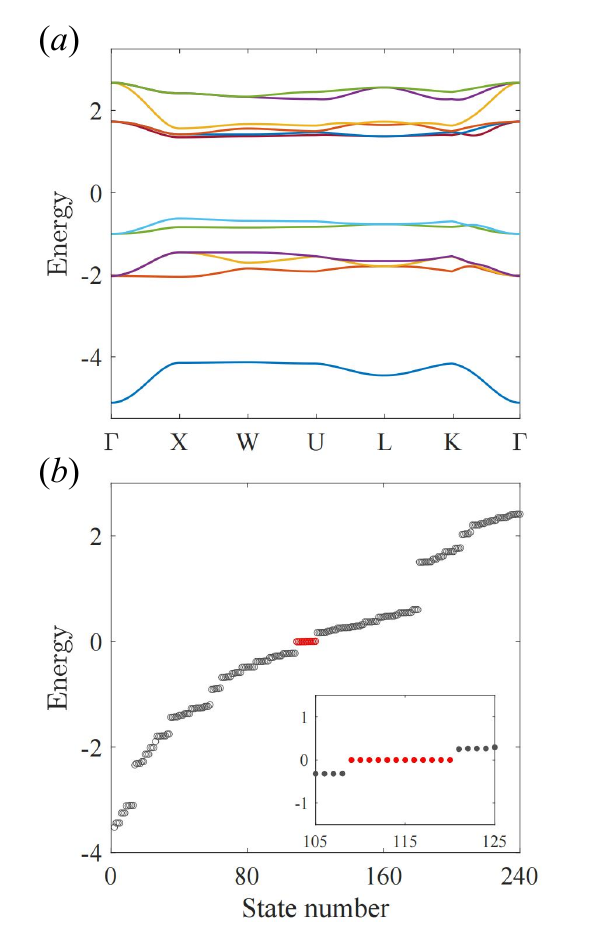}
\caption{(a) Band structures of $p$-orbital breathing pyrochlore lattice. (b) The energy spectrum of a four-layer ($L=4$) tetrahedron-shaped finite structure. In both subfigures, the hopping amplitudes are set to $t_{1\sigma}=-0.19, t_{2\sigma}=-1, t_{1\pi}=0, t_{2\pi}=0.18$. The 12 degenerate corner states are magnified in the inset of subfigure (b).}
\label{f4}
\end{figure}

\begin{equation}
\begin{aligned}
\bm{A}_{mn}(\bm{k})=i\langle u_m(\bm{k})|\partial_{\bm{k}}|u_n(\bm{k}) \rangle, \quad m,n=1,2,...,\nu, 
\end{aligned}
\end{equation}
where $|u_n(\bm{k}) \rangle$ is the periodic Bloch wave function for the $n$th bands, and $\nu$ is the number of bands under a band gap. For the lowest $\nu$ bands under a gap, the Berry phase is defined as
\begin{equation}
\begin{aligned}
\theta^{\nu}=\int_{P_i} Tr[\bm{A}(\bm{k})] d\bm{k},
\end{aligned}
\end{equation}
in which $P_i$ ($i=1,2,3,4$) represents an integral path ($W_i \to \Gamma \to W_{i+1}$ with $W_5=W_1$) in the Brillouin zone when using the Wilson-loop approach \cite{fukui2005chern,yu2011equivalent}, as shown in Fig. \ref{f3}(a). In fact, the $Z_4$ Berry phase in momentum space here is equivalent to the local gauge twists berry phase in parameter space discussed in Refs. \cite{kariyado2018z,otsuka2021higher}. Besides, the four high-symmetry points $W_1$, $W_2$, $W_3$, and $W_4$, shown in Fig. \ref{f3}(a), are equivalent because of the $S_4$ symmetry. 
Therefore, there are four equivalent integral paths ( $P_1: W_1 \to \Gamma \to W_2, P_2: W_2 \to \Gamma \to W_3, P_3: W_3 \to \Gamma \to W_4, P_4: W_4 \to \Gamma \to W_1$ ) leading to identical $\theta^{\nu}$, that is,
\begin{equation}
\begin{aligned}
\theta^{\nu}(P_1)=\theta^{\nu}(P_2)=\theta^{\nu}(P_3)=\theta^{\nu}(P_4).
\end{aligned}
\end{equation}
Obviously, the sum of the integrals along the four paths ($P_1$, $P_2$, $P_3$, and $P_4$) is zero:
\begin{equation}
\begin{aligned}
\sum_{i=1}^{4}\theta^{\nu}(P_i)=0.
\end{aligned}
\end{equation}
Therefore, the $Z_4$ Berry phase is quantized as
\begin{equation}
\begin{aligned}
\theta^{\nu}\equiv\theta^{\nu}(P_i)=2\pi n/4, \quad (n=0,1,2,3).
\end{aligned}
\end{equation}

We present the bulk topological phase diagram in Fig. \ref{f3}(b) and \ref{f3}(c), 
where different colors indicates the values of $Z_4$ berry phase.
It should be noted that the origin of topology is the adiabatic connection between the energy spectrum at $t_1\ne0$ and $t_1=0$ \cite{ezawa2018higher}. 
The $Z_4$ berry phase changes when the band gap is closed and reopened. 
Since the higher-order topological corner states appear at zero energy, we take into account an additional 
condition when plotting the topological phase diagram:
the zero energy states must exist inside a band gap (\textit{i.e.}, the bulk bands cannot cross the zero-energy level).
The topologically nontrivial regions in the phase diagram of Fig. \ref{f3}(b) and \ref{f3}(c) are actually smaller than those obtained by only considering the $Z_4$ berry phase. 

Figure \ref{f3}(b) illustrates the phase diagram in the $\alpha-\beta$ plane.
Two nontrivial $Z_4$ berry phases are identified: $\theta^3/2\pi=0.25$ and $\theta^6/2\pi=0.5$, calculated using the 
lowest three ($\nu=3$) and six ($\nu=6$) bands, respectively. These phases are represented by the orange and blue 
regions in the phase diagram. Due to the additional condition used, the two regions do not overlap 
but intersect at a point ($\alpha=0$, $\beta\approx0.33$).
Figure \ref{f3}(c) shows the phase diagram in the $t_{1\sigma}-t_{2\pi}$ plane, with the other two parameters being 
fixed at $t_{1\pi}=0$ and $t_{2\sigma}=-1$. This phase diagram corresponds to the simulated acoustic crystal scenario discussed in 
in Sec. \hyperlink{s5}{\uppercase\expandafter{\romannumeral5}}, where the radius of the intracell connecting cylinder in the acoustic crystal is small, making $t_{1\pi}$ negligibly small compared to $t_{1\sigma}$. 
In Fig. \ref{f3}(c), there is only one nontrivial $Z_4$ Berry phase, $\theta^6/2\pi=0.5$, calculated using the six bands 
under the zero-energy gap. The red dot in the nontrivial blue region indicates the parameter values 
corresponding to the acoustic simulation in Sec. \hyperlink{s5}{\uppercase\expandafter{\romannumeral5}}, 
which are $t_{1\sigma}=-0.19$, $t_{2\sigma}=-1$, $t_{1\pi}=0$, $t_{2\pi}=0.18$. 
The band structures for this set of parameter values are depicted in Fig. \ref{f4}(a) for a comparison with the acoustic simulation results.


\begin{figure}[t]
\centering
\includegraphics[width=0.99\linewidth]{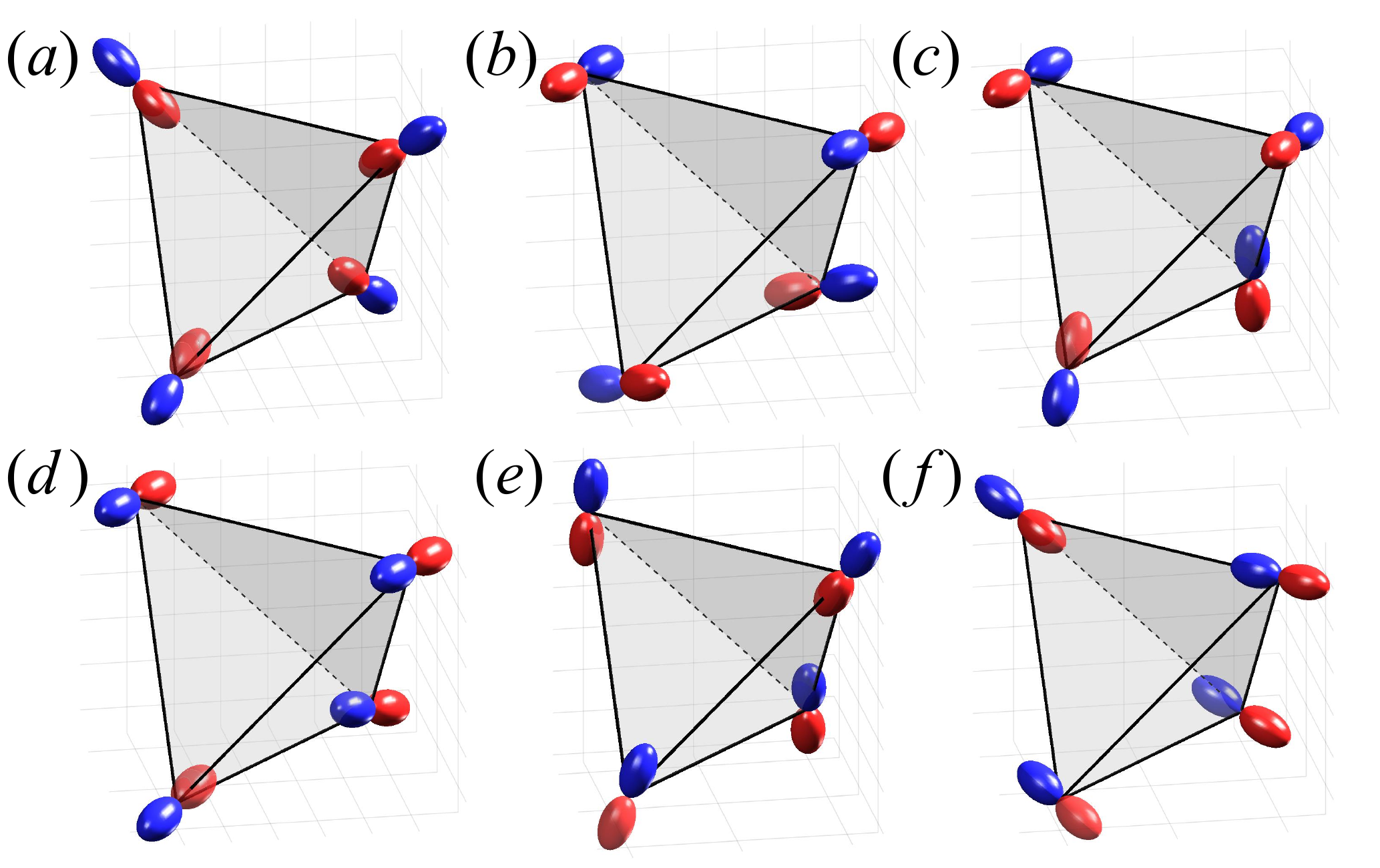}
\caption{The orbital configurations of corner states for a four-layer ($L=4$) tetrahedron-shaped finite structure. (a)-(f) Only the orbitals on the corner are shown. The dashed line represents an edge on the far side of a tetrahedron.
The direction of the major axis of the ellipse indicates the direction of the $p$ orbital. 
The parameters are chosen as $t_{1\sigma}=-0.19, t_{2\sigma}=-1, t_{1\pi}=0, t_{2\pi}=0.18$, 
which are the same as those in Fig. \ref{f4}.}
\label{f5}
\end{figure}

\begin{figure*}[t]
\centering
\includegraphics[width=0.99\linewidth]{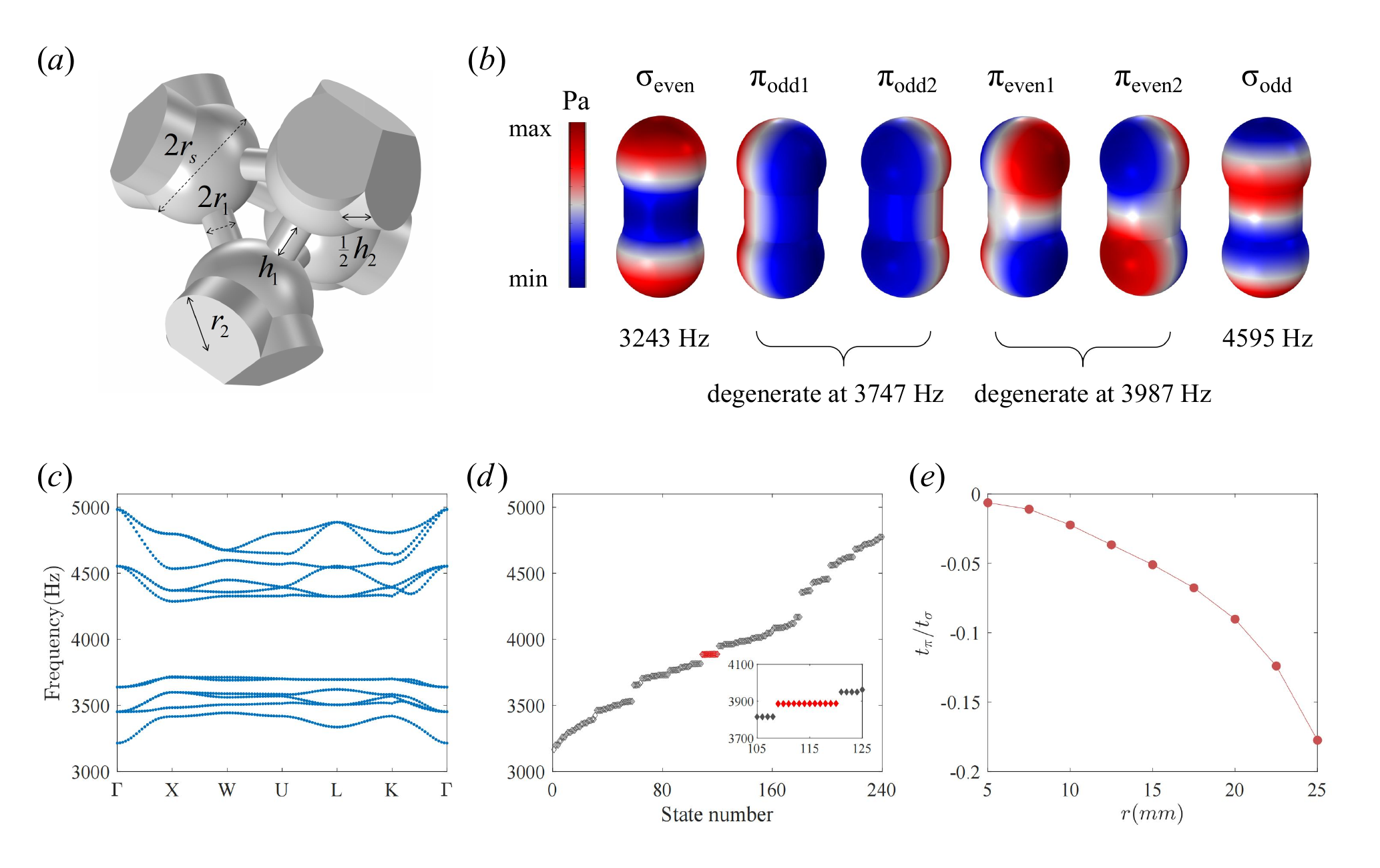}
\caption{(a) Schematic diagram for the unit cell of acoustic breathing pyrochlore lattices where $r_s=30$ mm, $r_1=7.6$ mm, $h_1=15.8$ mm, $r_2=25$ mm and $h_2=31.8$ mm. (b) The $p$-orbital pressure fields of two coupled acoustic resonators. The parameters are set to $r_s$, $r_2$ and $h_2$ in (a). (c) The simulated band structure. (d) Evolution of the spectrum (frequency) of four-layer ($L=4$) finite structure as a function of state numbers. (e) The ratio of $t_{\pi}$ and $t_{\sigma}$ hopping amplitude as a function of the radius of the cylindrical connector where $r_s$ and $h_2$ are fixed.}
\label{f6}
\end{figure*}

\hypertarget{s4}{}
\section*{\romannumeral4. The orbital corner states}

Here we investigate the $p$-orbital corner states in a finite tetrahedral structure for the breathing pyrochlore lattice. 
We use $L$, the number of unit cells (the small tetrahedron surrounded by red lines shown in Fig. \ref{f1}(a)) along one edge of the finite lattice, to define the size of our structure \cite{ezawa2018higher}. 
The total number of unit cells in the finite structure is $L(L + 1)(L+2)/6$, 
corresponding to $4L(L + 1)(L+2)/6$ sites and $12L(L + 1)(L+2)/6$ $p$-orbital states. 
We take $L=4$ in this work, indicating 20 unit cells, 80 sites, and 240 $p$-orbital states. 

\hypertarget{s4.1}{}
\subsection*{{\romannumeral4}.1 The parameter space of $\alpha$ and $\beta$}

The energy dispersion of a tetrahedron-shaped finite structure
is obtained by diagonalizing the real-space Hamiltonian in Eq. (\ref{q6}). 
We first discuss the corner states in the parameter space of $\alpha$ and $\beta$.
For the parameters $\alpha=0.1$, $\beta=0.2$, the $Z_4$ berry phase (mod $2\pi$) has 
a value of 0.5 [see Fig. \ref{f3}(b)], 
with six bands under the zero-energy gap in the bulk spectrum [see Fig. \ref{f2}(a)]. 
Whereas for $\alpha=-0.1$, $\beta=0.65$, the $Z_4$ berry phase (mod $2\pi$) has a value of 0.25, 
with three bands under the zero-energy gap [see Fig. \ref{f2}(d)]. 
We plot the energy as a function of the number of eigenstates in Fig. \ref{f2}(c) and \ref{f2}(f). 
Both of them exhibit 12 degenerate zero-energy corner states, as displayed in the enlarged insets. 
For these two $Z_4$ berry phases, the orbital configurations of corner states exhibit similar characteristics, 
since the energy spectra of both phases can be adiabatically connected to the same limit of $t_1=0$.


\begin{figure*}[t]
\centering
\includegraphics[width=0.99\linewidth]{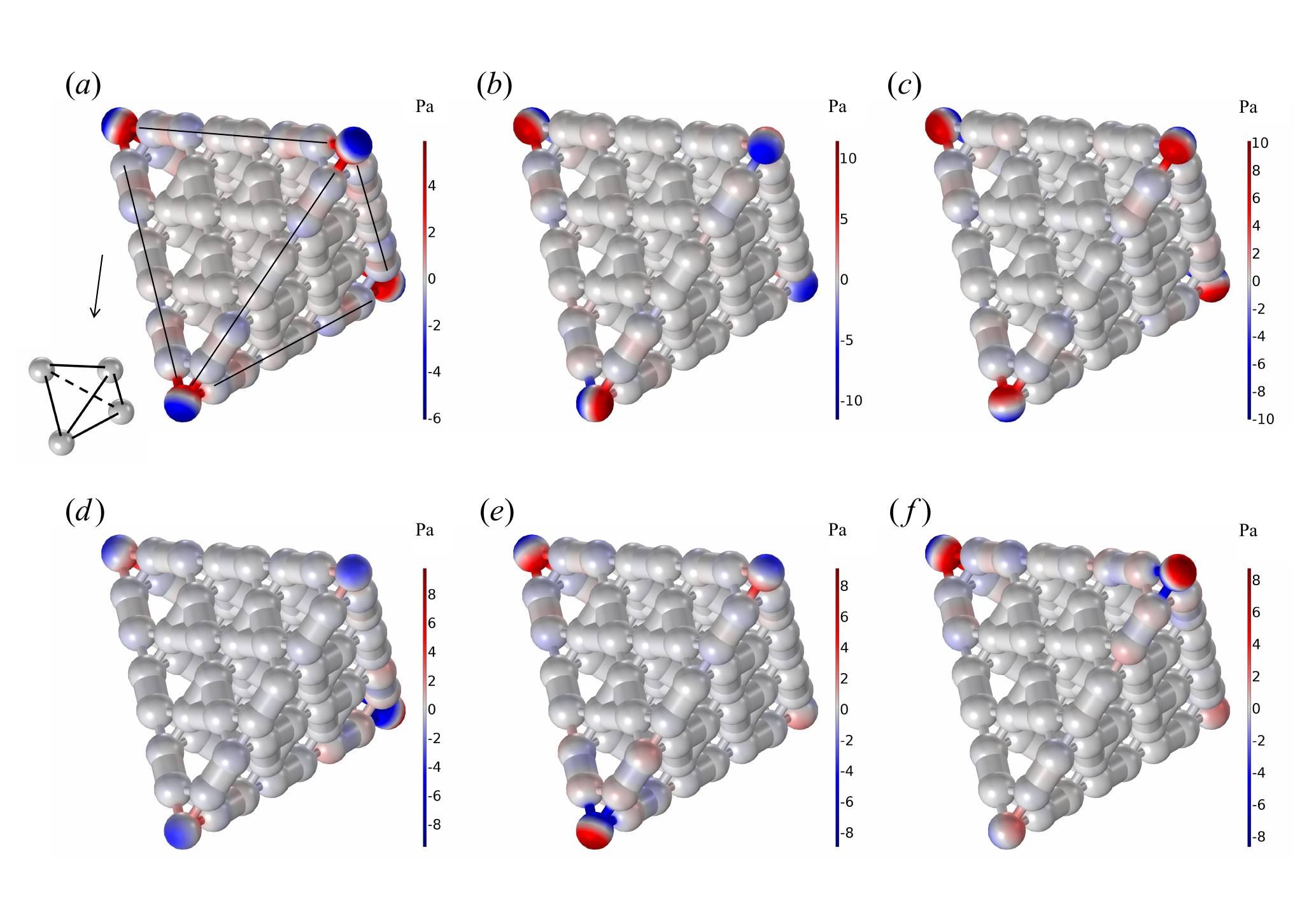}
\caption{(a)-(f) The $p$-orbital acoustic pressure fields of the tetrahedron made of breathing pyrochlore lattices with $L=4$.
In order to illustrate the 3D structure, we add a schematic tetrahedron in the lower-left corner of sub-figure (a) 
to guide the eye.}
\label{f7}
\end{figure*}

The range of parameters with distinguishable corner states may be smaller than that predicted by the bulk topological invariants (see Fig. \ref{f3}). This is because the localized states of the finite tetrahedral structure (existing on the edges, surfaces etc.) could have energies very close to zero, and therefore obscure the zero-energy corner states.
Clearly, the zero-energy states at the right and left end of the red line in Fig. \ref{f2}(b) and \ref{f2}(e) cannot 
be distinguished whether they are corner states or not. 
In addition, the appearance of corner states are strongly affected in two parameter regions around  
$\beta=0$ ($t_{\pi}=0$) and $\beta=0.33$, due to the possible mixing with bulk bands. 
As shown in Fig. \ref{f2}(b) and \ref{f2}(e), the zero-energy band (red line) is located in a gap surrounded by bulk bands.
When the value of $\beta$ is around zero, this gap will disappear, leading to the indistinguishable zero-energy states for corner. 
When $\beta$ is around 0.33, some bulk bands will rise up to the zero-energy position, and cover the corner states. This phenomenon ($\beta \approx 0.33$) can also be explained by the band structure obtained from the $k$-space Hamiltonian, that is, the fourth, fifth, and sixth bands (counting from the bottom) will gradually rise as $\beta$ increases from -1 to 1. These three bands just pass through the zero-energy level when $\beta$ is about $0.33$.


\hypertarget{s4.2}{}
\subsection*{{\romannumeral4}.2 The parameter space of $t_{1\sigma}$ and $t_{2\pi}$}

In the parameter space of $t_{1\sigma}-t_{2\pi}$, relevant to the acoustic model,
we focus on the following typical set of parameter values: 
$t_{1\sigma}=-0.19$, $t_{2\sigma}=-1$, $t_{1\pi}=0$, $t_{2\pi}=0.18$; see the red dot in Fig. \ref{f3}(c).
We plot in Fig. \ref{f4}(b) the energy spectrum of a four-layer ($L=4$) tetrahedron-shaped finite structure. 
There are 12 degenerate corner states within a sizable gap, 
similar to the situations discussed in Sec. \hyperlink{s4.1}{\uppercase\expandafter{\romannumeral4}.1}.
The orbital configurations of various corner states are displayed in Fig. \ref{f5} for this finite structure.
For clarity, we only exhibit the $p$-orbital orientations at four corners. 
One of the most symmetrical states appears in Fig. \ref{f5}(a). 
There are also three orbital configurations in which the orbitals at four corner sites are nearly parallel, as shown in Fig. \ref{f5}(d)-(f).
Part of the configurations in Fig. \ref{f5}(b) and \ref{f5}(c) exhibit "staggered $p$-orbital order".
Interestingly, the distributions of these configurations can be observed in an acoustic system; see the next section.

Again, the corner states within a sizable gap appear in a small region of the blue area in the bulk topological phase diagram in Fig. \ref{f3}(c), as the localized states of the finite structure emerge within the bulk band gap.
In Fig. \ref{f4}(b), in addition to the zero-energy corner states, three types of 
localized states are identified: surface states, edge states, and Type-II corner states.
Specifically, the type-II corner states, which exhibit profiles similar to topological corner states but decay exponentially away from the corners \cite{li2020higher,gao2024acoustic}, 
also appear in this 3D $p$-orbital system. 
As shown in Fig. \ref{f4}(b), the gap harboring the corner states closes as $t_{2\pi}$ approaches zero, 
highlighting the crucial role of $t_{2\pi}$ in the $p$-orbital system for observing the corner states.

\hypertarget{s5}{}
\section*{\romannumeral5. The realization of orbital corner states in acoustic crystals}
\label{simu}

Recently, topological acoustics have been a large research area to mimic topological phenomena in condensed matter system \cite{yang2015topological,ma2019topological,hu2021non,xue2022topological,zhu2023topological}. Compared to other systems, topological acoustic crystals have excellent tunability which can be more easily designed into various geometric shapes to achieve corresponding topological properties. At the same time, acoustic properties can be accurately measured in a wide range of laboratory conditions, allowing for direct and convenient acquisition of topological information. Therefore, we use acoustic system to realize our tight-binding model. Instead of using a conventional cylindrical resonators \cite{xue2019realization,weiner2020demonstration}, spherical resonators  \cite{chen2022observation,cheng2024three} are used to achieve a uniform space for $p$-orbital couplings.

We perform simulation with finite-element method to calculate the eigenstates of such acoustic breathing pyrochlore lattices, where sound hard boundary condition of all air-solid interfaces is considered. The density of the air and the speed of sound are set to be 1.25 kg/m$^{3}$ and 343 m/s. Unit cell of pyrochlore lattice is shown in Fig. \ref{f6}(a), the radius $r_s$ of sphere is fixed at 30 mm, the length and radius of the connecting cylinders are used to adjust the strength of different hoppings where one type ($t_1$) is set with radius $r_1$ of 7.6 mm and length $h_1$ of 15.8 mm, the other type ($t_2$) is 25 mm for the radius $r_2$ and 31.8 mm for the length $h_2$. 
In simulation, we slightly reduce the radius (to 29.4mm) of the spherical resonator at four corners, since the effective capacity of the resonator at corner is a bit larger than that in bulk if we use the same radius for all spheres.

In order to introduce $\pi$ hoppings in this acoustic model, here we consider two coupled spherical resonators connecting by a cylindrical tube whose parameters are set to $r_s$, $r_2$ and $h_2$. The $p$-orbital eigenfields are shown in Fig. \ref{f6}(b) where we can see $\sigma$-like and $\pi$-like bonding with even or odd parity in such coupled resonators. The hopping amplitudes $t_{\sigma}$ and $t_{\pi}$ in the acoustic model can be determined by half of the difference between the frequencies of the even and odd hybrid modes of two coupled resonators, i.e., $t_{\sigma}\propto 1/2(f_{\sigma_{odd}} - f_{\sigma_{even}})$, $t_{\pi}\propto 1/2(f_{\pi_{odd}} - f_{\pi_{even}})$ \cite{chen2022observation}. 
In Fig. \ref{f6}(e), the ratio of estimated $t_{\pi}$ and $t_{\sigma}$ is plotted as a function of 
the radius of the connecting cylinder. As illustrated, a larger cylinder radius introduces a considerable $\pi$ hopping amplitude, which can open a gap to harbor the corner states.

The $p$-orbital band structure, shown in Fig. \ref{f6}(c), shows a band gap existed in the frequency range from 3700 Hz to 4300 Hz. Besides, the energy spectrum for a 4-layer finite structure is calculated in Fig. \ref{f6}(d) where 12 degenerate corner states appear in the $p$-orbital band gap. 
The acoustic band structure and energy spectrum of the finite structure qualitatively agree with those of the tight-binding model presented in Fig. \ref{f4}(a) and \ref{f4}(b), where the parameters are estimated from the double-resonator model. The deviation from the tight-binding model can be attributed to the fact that the coupling strength in a three-dimensional acoustic lattice differs from that in the double-resonator model due to the specific interconnections.

Starting from the acoustic parameters in Fig. \ref{f6}, if we decrease the radius $r_2$ of the cylinder, 
the corresponding $t_{2\pi}$ and $t_{2\sigma}$ both decrease.
However, $t_{2\pi}$ decreases much faster, i.e., the ratio $t_{2\pi}/t_{2\sigma}$ approaches zero as 
$r_2$ decreases; see Fig. \ref{f6}(e). 
In this way, the gap harboring the corner states will decrease [see the inset in Fig. \ref{f6}(d)], causing the corner states to mix with other localized states.
If we increase $r_2$ such that it approaches $r_s$, the tight-binding approximation for the $p$-band model becomes invalid, causing the corner states to disappear. When the cylinder radius $r_1$ is increased, the topological nontrivial region in Fig. \ref{f3}(c) shrinks, also making the corner states difficult to observe.

The acoustic fields of corner states are further displayed in Fig. \ref{f7}(a)-(f). We can see that the fields are highly localized on the four corners with different orbital configurations. In detail, Fig. \ref{f7}(a) displays the most symmetric geometry of $p$-orbitals pointing towards the center of the tetrahedron, which is consistent with the tight-binding results in Fig. \ref{f5}(a). Parallel states [see Fig. \ref{f7}(d)-(f)] exhibit a tendency for the orbital orientation of all four corners to align in one direction. The results of $p$-orbital acoustic field patterns are in qualitative agreement with tight-binding model shown in Fig. \ref{f5}.

\hypertarget{s6}{}
\section*{\romannumeral6. Conclusion}
In summary, we have investigated the third-order topological orbital corner states in the $p$-orbital breathing pyrochlore lattice. 
The presence of two orthogonal $\pi$-type hopping terms is the crucial factor to obtain the orbital corner states. Without them, the corner states are indistinguishable from the bulk states.
Based on the $Z_4$ berry phase, we investigate the bulk topological properties and obtain the phase diagram. 
By constructing a four-layer regular tetrahedral finite structure, we demonstrate the existence of corner states and reveal 
their rich orbital configurations.
Finally, we realize the orbital corner states in acoustic crystals and validate the experimental feasibility of the $p$-orbital model.
Our work may further promote the research of high-order topology with orbital freedom.

\section*{ACKNOWLEDGMENT}
We are grateful for the discussions with Chuanshu Xu, Peng Long and Qiang Wei. We also thank Huanyang Chen for the insightful suggestions.
This work is supported by the National Natural Science Foundation of China (Grants No. 11974293) and the Natural Science Foundation of Fujian Province (Grant No.2024J01081).

\bibliographystyle{apsrev4-1}

\bibliography{reference}

\end{document}